\documentclass[aps,prd,nofootinbib,showpacs]{revtex4}
\usepackage[utf8x]{inputenc}
\usepackage{fancyhdr}
\usepackage{graphics}
\usepackage{epsfig}
\usepackage{graphicx}
\usepackage{epsf}
\usepackage{enumitem}
\usepackage{amsmath, amsthm, amssymb}
\usepackage{slashed}
\usepackage{subfig}
\usepackage{nicefrac}
\usepackage{url}
\usepackage{color}
\usepackage{appendix}

\newcommand{\pop}{\phi_1^+}
\newcommand{\pom}{\phi_1^-}
\newcommand{\ptp}{\phi_2^+}
\newcommand{\ptm}{\phi_2^-}
\newcommand{\poz}{\phi_1^0}
\newcommand{\ptz}{\phi_2^0}
\newcommand{\re}{{\rm Re}(\mu_3^2)}

\begin{document}
\title{Perturbative Analysis of the Electron EDM and CP Violation in Two Higgs Doublet Models}
\author{Seyda Ipek}
\affiliation{Department of Physics, University of Washington, Seattle, WA 98195, USA}
\date{\today}

\begin{abstract}
I consider a general two Higgs doublet model with CP violation. I give a perturbative expansion for the mass eigenstates in terms of the small CP violating phase. I use these analytical expressions to show that $\mathcal{O}(10^{-2})$ CP violation is allowed by the experimental bounds on the electron electric dipole momnent in some regions of the parameter space. These regions also include parameters that are expected to give a strongly first order electroweak phase transition required for electroweak baryogenesis. I also comment on how to incorporate the CP violation into the searches for a strongly first order electroweak phase transition which could explain the matter/anti-matter asymmetry in the universe.  
\end{abstract}

\maketitle

\section{Introduction}\label{intro}
In the universe there is far more matter than antimatter. The observed amount of baryon number ($B$) requires an asymmetry between quarks and antiquarks in the early universe of one part in $10^{-8}$. Sakharov formulated three necessary and sufficient rules in order for the universe to produce such a matter/antimatter discrepancy. These \textit{Sakharov conditions} for baryogenesis are: (1) $B$ is not conserved. (2) C and CP are not conserved. (3) The universe should not be in thermal equilibrium. The first two rules are obvious to produce more baryons over antibaryons. The third rule is required because in thermal equilibrium the number of baryons and antibaryons are equal in a CPT conserving theory. Hence as long as the universe is in equilibrium, the \emph{net} baryons that are produced will be washed out. 

It was first suggested by Kuzmin, Rubakov and Shaposhnikov \cite{Kuzmin198536} that baryogenesis can occur via an electroweak (EW) phase transition. The process is then called \textit{electroweak baryogenesis} (EWBG). A quick check of the Sakharov conditions shows that: (i) $B$ is not conserved in the Standard Model (SM) due to the rapid anomalous $B$ violation at high temperatures. (ii) There is CP violation in the electroweak sector of the SM coming from the Cabibbo-Kobayashi-Maskawa (CKM) matrix. (iii) If the EW phase transition is first order, it can accommodate the out-of-equilibrium conditions via bubble nucleation. These are the basic ingredients of EWBG; an extensive review can be found in \cite{Cohen:1993nk}. EWBG has the following potential problems:

\begin{enumerate}
 \item The interactions following the production of a net baryon number should conserve baryon number. Otherwise the produced baryons would be washed out. EW interactions violate $B$ via a $U(1)$ anomaly. Even though the rate of $B$ violation is exponentially small ($e^{-4\pi/\alpha_w} \sim 0$) at the weak scale (since it happens via a tunneling process between vacua with different baryon numbers), it is very rapid at high temperatures ($T > 100$ GeV), and would wash out all the baryons produced, for example, at the GUT scale. However, there is a conserved quantum number, baryon $-$ lepton number ($B-L$) in the SM that can be the source of a net baryon number production. This idea works even if neutrinos are Majorana particles (and hence $L$ is not conserved). In the case of Majorana neutrinos, the baryogenesis scale puts an upper bound on the neutrino masses \cite{nelson_upper_1990}. 
 
 \item CP violation in the SM comes from the CKM matrix. One can use the Jarlskog invariant \cite{jarlskog_commutator_1985} to give a parametrization invariant measure of CP violation. The Jarlskog invariant is defined as
 \begin{align}
  J &= {\rm Im}(V_{ij}V_{kl}V^*_{kj}V^*_{il}) \\
    &= c_{12}c_{23}c_{13}^2s_{12}s_{23}s_{13}\sin\delta \hspace{.5in} (\text{in Kobayashi - Maskawa parametrization})
 \end{align}
where $c_{ij} \equiv \cos\theta_{ij}$ is the mixing angle between $i$ and $j$ quark, and $\delta$ is the phase in the CKM matrix. Notice that even if the CP violating phase is large, it gets multiplied by small mixing angles, making $J$ $(\sim 10^{-5})$ small. Furthermore, since CP violating processes involve all three generations of quarks, they are suppressed by small Yukawa couplings. It has been shown that the CP violation in the SM results in a $B$ asymmetry of $10^{-20}$ \cite{gavela_standard_1994, huet_electroweak_1994}. Hence, we need to extend the SM to allow more CP violation. 

 \item If the EW phase transition is first order, the transition causes bubble nucleation. A first order phase transition needs the order parameter, in this case the expectation value of the Higgs filed, $v(T)$, to change abruptly at the critical temperature ($T_c\sim 100$ GeV). Inside the bubbles is the broken phase, with a non-zero $v$ of the Higgs field. The rate of $B$ violation inside the bubble is proportional to $e^{-4\pi v/g_2T}$, where $g_2$ is the $SU(2)$ coupling constant and $T$ is the temperature immediately after the phase transition \cite{Kuzmin198536}. This phase transition can only produce the required non-equilibrium conditions if the rate of $B$ violation is slower than the expansion rate of the universe. This puts a bound on the Higgs expectation value ($v(T)/T \gtrsim 1$) and consequently on the Higgs mass. A strongly first order EW phase transition requires a Higgs of mass less than 60 GeV in the SM. Alternately, with a Higgs mass of 125 GeV, the transition is a crossover. Rapid $B$ 
violation would then let the baryon number equilibrate to zero if $B - L$ is zero. One solution to this problem is to extend the scalar sector of the SM to have two Higgs doublets. 
 
\end{enumerate} 

These are the reasons why we will study two Higgs doublet models (2HDMs). It is known that in a 2HDM, one can have a first order electroweak phase transition \cite{bochkarev_model_1991, Dorsch:2013wja}. In addition, one can have CP violation in this extended Higgs sector to generate enough CP violation for baryogenesis. Recently Ref \cite{Dorsch:2013wja} did a numerical analysis of the phase space and showed that CP conserving 2HDMs with a Higgs boson of mass 125 GeV can indeed accommodate a strongly first order phase transition. The literature for 2HDMs is mostly based on CP conserving models because the analysis is simpler. This is because in the case of CP violation, the CP-even and CP-odd fields mix, making the definition of mass eigenstates more complex. This complexity limited the studies of 2HDMs to either CP conserving models or to small portions of the parameter space of CP violating models, such as those with large mass splittings between the Higgs bosons or specific ordering of the masses (e.g. \
\cite{shu_impact_2013}). A more 
general study of the CP violating 2HDM is required especially for baryogenesis purposes. In this work, I use the fact that we expect the new CP violating phase to be of $\mathcal{O}(0.1)$ or less, and give a perturbative expansion of the Higgs states in terms of a small CP violating phase. I focus on the portion of the parameter space that gives a first order phase transition (e.g. $\tan\beta<4, \mu_3>100$ GeV, $m_{\tilde{A}}>300$ GeV), and I find the allowed CP violation that is consistent with experimental limits on electron electric dipole moment (EDM). I also comment on how to include CP violation in searches for a first order phase transition in an easy way. 

The rest of this paper is organized as follows: In Section \ref{2hdm}, I give the general form of the scalar potential for 2HDMs, and calculate the approximate mass eigenstates for the CP violating case. In Section \ref{2hdmpt}, I show how to include CP violation in phase transition analyses of the EWBG. In Section \ref{edmsec}, I show how to constrain the CP violating 2HDM from the limits on the electron EDM. In Section \ref{conc}, I give my concluding remarks.

\section{the two Higgs Doublet Model}\label{2hdm}
In this section, I briefly introduce 2HDM, an extension of the SM. Our motivation, as discussed in the Introduction, is to accommodate EWBG\footnote{Of course, apart from baryogenesis reasons, two Higgs doublets are a necessary feature of supersymmetry.}. The model has two scalar $SU(2)$ doublets, $\Phi_1$ and $\Phi_2$. Together, they have eight degrees of freedom, three of which are fictitious Nambu-Goldstone bosons that are eaten by the gauge bosons. Hence there are five physical Higgs bosons left: two charged, and three neutral. If there is no CP violation, two of the three neutral Higgs bosons are CP-even, and the third one is CP-odd. In the case of CP violation, the CP eigenstates are not the mass eigenstates, since the CP-odd and CP-even Higgs fields mix.

One can impose a $Z_2$ symmetry with $\Phi_1 \rightarrow \Phi_1$ and $\Phi_2 \rightarrow -\Phi_2$, so that there are no flavor changing neutral currents (FCNCs) at tree level\footnote{There are other ways to suppress FCNC in the two Higgs doublet model, e.g. \cite{a2hdm}.}. Generically, one can say $\Phi_2$ always couples to up-type quarks. The rest of the fermion couplings are such that $\Phi_1$ and $\Phi_2$ do not both couple to fermions of a given charge. Hence, there are four types of 2HDMs that do not induce FCNCs. Later, we will see that this $Z_2$ symmetry can be softly broken, which will be the source of the CP violation in the scalar sector. For a more complete treatment of 2HDMs, see the review in \cite{Branco:2011iw}.  The top quark coupling is by far the most important Yukawa coupling for EWBG. Therefore, phase transition analysis for baryogenesis purposes can be done without specifying a particular 2HDM. 

The most general potential that has a softly broken $Z_2$ symmetry can be written as \cite{hhg}:
\begin{align}\label{2hp}
 V = \lambda_1&\left(\Phi_1^\dagger\Phi_1-\frac{v_1^2}{2}\right)^2+\lambda_2\left(\Phi_2^\dagger\Phi_2-\frac{v_2^2}{2}\right)^2+\lambda_3\,\left[\left(\Phi_1^\dagger\Phi_1-\frac{v_1^2}{2}\right)+\left(\Phi_2^\dagger\Phi_2-\frac{v_2^2}{2}\right)\right]^2 \nonumber \\
 &+\lambda_4\,\biggl[(\Phi_1^\dagger\Phi_1)(\Phi_2^\dagger\Phi_2)-(\Phi_1^\dagger\Phi_2)(\Phi_2^\dagger\Phi_1)\biggr]+\lambda_5\,\left({\rm Re}(\Phi_1^\dagger\Phi_2)-\frac{v_1v_2}{2}\,\cos\xi\right)^2+\lambda_6\,\left({\rm Im}(\Phi_1^\dagger\Phi_2)-\frac{v_1v_2}{2}\sin\xi\right)^2
\end{align}
where $\lambda_i$ are real. The minimum of the potential is at:
\begin{align}
 \langle\Phi_1\rangle = \frac{1}{\sqrt{2}}\left(
\begin{array}{c}
0\\
v_1\\
\end{array}
\right)    \hspace{0.5in}  \langle\Phi_2\rangle = \frac{1}{\sqrt{2}}\left(
\begin{array}{c}
0\\
v_2 e^{i\xi}\\
\end{array}
\right)
\end{align}
The angle $\xi$ is the CP violating phase. If $\lambda_5=\lambda_6$, we can rotate away this phase, and there is no CP violation.

\subsection{CP Conserving 2HDM} \label{cpcons2hdm}
Let us start with the CP conserving 2HDM ($\xi=0$). The two Higgs doublets can be written as:

\begin{align} \label{2hds}
 \Phi_1 = \frac{1}{\sqrt{2}}\left(
\begin{array}{c}
\pop\\
v_1+\poz\\
\end{array}
\right)    \hspace{0.5in}  \Phi_2 =\frac{1}{\sqrt{2}} \left(
\begin{array}{c}
\ptp\\
v_2+\ptz\\
\end{array}
\right)
\end{align}

The mass eigenstates are found by inserting $\Phi_1$ and $\Phi_2$ into Eqn.\ref{2hp}. The charged Higgs fields are:
\begin{align}
 H^\pm =\frac{1}{\sqrt{2}} ( -\sin\beta\,\phi_1^\pm+\cos\beta\,\phi_2^\pm )
\end{align}
with mass $m_{H^\pm}^2=\frac{\lambda_4}{2}\,v^2$, where $\tan\beta=\frac{v_2}{v_1}$ and $v^2=v_1^2+v_2^2$. Here $v=246$ GeV, which is set by the $W/Z$ mass. The two fields that are orthogonal to $H^\pm$ are the the two fictitious Nambu-Goldstone bosons:
\begin{align}
  G^\pm = \frac{1}{\sqrt{2}}(\cos\beta\,\phi_1^\pm+\sin\beta\,\phi_2^\pm)
\end{align}

The massive CP-odd Higgs field is:
\begin{align}
 A^0=-\sin\beta\,{\rm Im}\poz+\cos\beta\,{\rm Im}\ptz
\end{align}
with mass $m_{A^0}^2=\frac{\lambda_6}{2}\,v^2$. The field that is orthogonal to $A^0$ is the third fictitious Nambu-Goldstone boson:
\begin{align}
 G^0=\cos\beta\,{\rm Im}\poz+\sin\beta\,{\rm Im}\ptz
\end{align}

The two CP-even states are the eigenstates of the following mass matrix (in $({\rm Re}\poz, \,\,\, {\rm Re}\ptz)$ space):
\begin{align} \label{mnocp}
 M^2= \left( \begin{array}{cc}
2v_1^2(\lambda_1+\lambda_3)+v_2^2\frac{\lambda_5}{2} & v_1v_2\left(2\lambda_3+\frac{\lambda_5}{2}\right)  \\
v_1v_2\left(2\lambda_3+\frac{\lambda_5}{2}\right) & 2v_2^2(\lambda_2+\lambda_3)+v_1^2\frac{\lambda_5}{2}
 \end{array} \right)
\end{align}
Its eigenvectors, $H^0$ and $h^0$, are:
\begin{align}
 H^0&= \cos\alpha\,{\rm Re}\poz+\sin\alpha\,{\rm Re}\ptz \\
 h^0&= -\sin\alpha\,{\rm Re}\poz+\cos\alpha\,{\rm Re}\ptz
\end{align}
with the eigenvalues:
\begin{subequations}\label{cpcmass}
\begin{align}
 m_{H^0}^2&=\frac{v^2}{4}\left(4\lambda_3+\lambda_5+4(\lambda_2\,s^2_\beta+\lambda_1\,c^2_\beta)+\sqrt{(4\lambda_3+\lambda_5)s^2_{2\beta}+[(-4\lambda_3+\lambda_5)c_{2\beta}+4(\lambda_2\,s^2_\beta-\lambda_1\,c^2_\beta)]^2}\right) \\
 m_{h^0}^2&=\frac{v^2}{4}\left(4\lambda_3+\lambda_5+4(\lambda_2\,s^2_\beta+\lambda_1\,c^2_\beta)-\sqrt{(4\lambda_3+\lambda_5)s^2_{2\beta}+[(-4\lambda_3+\lambda_5)c_{2\beta}+4(\lambda_2\,s^2_\beta-\lambda_1\,c^2_\beta)]^2}\right)
\end{align}
\end{subequations}
where $s_\beta\equiv\sin\beta$. Notice that $h^0$ is the lighter Higgs. In the rest of this paper, we will take $h^0$ to be the observed Higgs, setting its mass, $m_{h^0}$, to 125 GeV. The angle $\alpha$ is the mixing angle between the real parts of the two Higgs fields, and it is defined as:
\begin{align} \label{cosa}
 \cos^2\alpha={1\over2}+\frac{(4\lambda_3+\lambda_5)c_{2\beta}+4(\lambda_2\,s^2_\beta-\lambda_1\,c^2_\beta)}{2\sqrt{(4\lambda_3+\lambda_5)s^2_{2\beta}+[(-4\lambda_3+\lambda_5)c_{2\beta}+4(\lambda_2\,s^2_\beta-\lambda_1\,c^2_\beta)]^2}}
\end{align}

\subsection{CP Violating 2HDM}\label{2hdmcpv}
In this section we will work with a CP violating 2HDM, and find the approximate mass eigenstates in terms of a small CP violating phase. The scalar potential is given in Eqn.\ref{2hp}, with the condition that $\lambda_5\neq\lambda_6$. The two Higgs doublets can be written as:
\begin{align}
 \Phi_1 = \frac{1}{\sqrt{2}}\left(
\begin{array}{c}
\pop\\
v_1+\poz\\
\end{array}
\right)    \hspace{0.5in}  \Phi_2 = \frac{1}{\sqrt{2}} \left(
\begin{array}{c}
\ptp\\
v_2e^{i\xi}+\ptz\\
\end{array}
\right)
\end{align}
Inserting these fields into the potential in Eq.\ref{2hp}, we get:
\begin{align}\label{2hcp}
 V = &(\lambda_1+\lambda_3)v_1^2({\rm Re}\poz)^2+(\lambda_2+\lambda_3)v_2^2({\rm Re}(e^{-i\xi}\ptz))^2+2\lambda_3v_1v_2\,{\rm Re}\poz\,{\rm Re}(e^{-i\xi}\ptz) \nonumber \\
 &+\frac{\lambda_4}{2}\biggl[v_1^2\ptp\ptm+v_2^2\pop\pom-2v_2v_2{\rm Re}(e^{-i\xi}\pop\ptm)\biggr] +\frac{\lambda_5}{2}\biggl[v_1^2({\rm Re}\ptz)^2+v_2^2({\rm Re}(e^{-i\xi}\poz))^2+2v_1v_2\,{\rm Re}\ptz\,{\rm Re}(e^{-i\xi}\poz)\biggr] \nonumber \\
 &+\frac{\lambda_6}{2}\biggl[v_1^2({\rm Im}\ptz)^2+v_2^2({\rm Im}(e^{-i\xi}\poz))^2 - 2v_1v_2\,{\rm Im}\ptz\,{\rm Im}(e^{-i\xi}\poz)\biggr] + \cdots
\end{align}
where the ellipses are for terms that are third order and higher. 

The charged massive Higgs fields are easy to read off from this potential:
\begin{align}
 H^{\pm}=\frac{1}{\sqrt{2}}( -\sin\beta\,\phi_1^\pm+\cos\beta\,(e^{-i\xi}\phi_2)^\pm)
\end{align}
with mass $m_{H^\pm}^2=\frac{\lambda_4}{2}\,v^2$. The two fictitious Nambu-Goldstone bosons are the ones orthogonal to $H^\pm$. 

As mentioned earlier, in the CP violating 2HDM, there is no mass eigenstates with definite CP. The CP-odd and CP-even components of the two Higgs fields mix, making the analysis more complicated. Here we will assume small CP violation and make a perturbative expansion of the three neutral states. The neutral part of the potential in Eqn.\ref{2hcp} can be written as:
\begin{align}
 V_{neutral}=\frac{1}{2} \biggl({\rm Im}\poz\,\,\,\,{\rm Im}\ptz\,\,\,\,{\rm Re}\poz\,\,\,\,{\rm Re}\ptz\biggr)\,\mathcal{M}^2\,\left(\begin{array}{c}
{\rm Im}\poz\\
{\rm Im}\ptz\\
{\rm Re}\poz\\
{\rm Re}\ptz
\end{array}\right)
\end{align}
where the mass matrix $\mathcal{M}$ is:
\begin{align}
 \mathcal{M}^2=v^2\left(\begin{array}{cccc}
                             \frac{1}{2}(\lambda_5s_\xi^2+\lambda_6c_\xi^2)s_\beta^2 & -\frac{\lambda_6}{2}s_\beta c_\beta c_\xi & \frac{1}{2}(\lambda_5-\lambda_6)s_\beta^2 s_\xi c_\xi & \frac{\lambda_5}{2}s_\beta c_\beta s_\xi \\
                             -\frac{\lambda_6}{2}s_\beta c_\beta c_\xi & 2(\lambda_2+\lambda_3)s_\beta^2 s_\xi^2+\frac{\lambda_6}{2}c_\beta^2 & \left(2\lambda_3+\frac{\lambda_6}{2}\right)s_\beta c_\beta s_\xi & 2(\lambda_2+\lambda_3)s_\beta^2s_\xi c_\xi  \\
                             \frac{1}{2}(\lambda_5-\lambda_6)s_\beta^2s_\xi c_\xi & \left(2\lambda_3+\frac{\lambda_6}{2}\right)s_\beta c_\beta s_\xi & 2(\lambda_1+\lambda_3)c_\beta^2 + \frac{1}{2}(\lambda_5c_\xi^2+\lambda_6s_\xi^2)s_\beta^2 & \left(2\lambda_3+\frac{\lambda_5}{2}\right)s_\beta c_\beta c_\xi \\
                             \frac{\lambda_5}{2}s_\beta c_\beta s_\xi & 2(\lambda_2+\lambda_3)s_\beta^2s_\xi c_\xi & \left(2\lambda_3+\frac{\lambda_5}{2}\right)s_\beta c_\beta c_\xi & 2(\lambda_2+\lambda_3)s_\beta^2 c_\xi^2 + \frac{\lambda_5}{2}c_\beta^2
                            \end{array}\right) \label{matrix}
\end{align}
This matrix has an eigenvector with a zero eigenvalue corresponding to the fictitious Nambu-Goldstone boson:
\begin{align}
 \tilde{G}^0=\cos\beta\, {\rm Im}\poz + \sin\beta\cos\xi\,{\rm Im}\ptz - \sin\beta\sin\xi\,{\rm Re}\ptz
\end{align}
We can rotate away the part of $\mathcal{M}^2$ that corresponds to the Nambu-Goldstone boson, hence reducing it to a $3\times3$ matrix. Then, we assume small $\xi$ and expand this $3\times3$ matrix up to $\mathcal{O}(\xi^3)$ (for details, see Appendix \ref{pert}):
\begin{align}\label{mex}
  \widetilde{M}^2=v^2\left(\begin{array}{ccc}
                   \frac{\lambda_6}{2}\textcolor{red}{-\frac{1}{2}(\lambda_6-\lambda_5)\xi^2} & \textcolor{blue}{\frac{1}{2}(\lambda_6-\lambda_5)s_\beta\,\xi} & \textcolor{blue}{\frac{1}{2}(\lambda_6-\lambda_5)c_\beta\,\xi} \\
                   \textcolor{blue}{\frac{1}{2}(\lambda_6-\lambda_5)s_\beta\,\xi} & 2(\lambda_1+\lambda_3)c_\beta^2+\frac{\lambda_5}{2}\,s_\beta^2\textcolor{red}{+\frac{1}{2}(\lambda_6-\lambda_5)\,s_\beta^2\xi^2} & \left(2\lambda_3+\frac{\lambda_5}{2}\right)\,s_\beta\,c_\beta\textcolor{red}{+\frac{1}{2}(\lambda_6-\lambda_5)\,s_\beta\,c_\beta\,\xi^2} \\
                   \textcolor{blue}{\frac{1}{2}(\lambda_6-\lambda_5)c_\beta\,\xi} & \left(2\lambda_3+\frac{\lambda_5}{2}\right)\,s_\beta\,c_\beta\textcolor{red}{+\frac{1}{2}(\lambda_6-\lambda_5)\,s_\beta\,c_\beta\,\xi^2} & 2(\lambda_2+\lambda_3)s_\beta^2+\frac{\lambda_5}{2}\,c_\beta^2\textcolor{red}{+\frac{1}{2}(\lambda_6-\lambda_5)\,c_\beta^2\xi^2}
                  \end{array} \right)
\end{align}
We can further simplify this matrix by diagonalizing the $2\times2$ bottom-right corner:
\begin{align}\label{mexd}
  \widetilde{M}^2=v^2\left(\begin{array}{ccc}
                   \frac{\lambda_6}{2}\textcolor{red}{-\frac{1}{2}(\lambda_6-\lambda_5)\xi^2} & \textcolor{blue}{\frac{1}{2}(\lambda_6-\lambda_5)s_\beta\,\xi} & \textcolor{blue}{\frac{1}{2}(\lambda_6-\lambda_5)c_\beta\,\xi} \\
                   \textcolor{blue}{\frac{1}{2}(\lambda_6-\lambda_5)s_\beta\,\xi} & \lambda_H\textcolor{red}{+\epsilon_H\xi^2} & 0 \\
                   \textcolor{blue}{\frac{1}{2}(\lambda_6-\lambda_5)c_\beta\,\xi} & 0 & \lambda_h\textcolor{red}{+\epsilon_h\xi^2}
                  \end{array} \right)
\end{align}
in the (approximate) basis $(A^0,\,\, H^0,\,\, h^0)$. Note that we are using the states $A^0,H^0,$ and $h^0$ from Section \ref{cpcons2hdm}. This is only for convenience, and they do not correspond to physical states. Similarly, we take $m_{H^0,h^0}^2 = \lambda_{H,h}v^2$, and
\begin{align*}
  \epsilon_H&= \frac{1}{2}(\lambda_6-\lambda_5)\sin^2(\alpha+\beta) \\
  \epsilon_h&=\frac{1}{2}(\lambda_6-\lambda_5)\cos^2(\alpha+\beta)
\end{align*}
where $\sin\alpha$ is the same as before up to an $\mathcal{O}(\xi^2)$ correction. As can be seen, when $\lambda_6=\lambda_5$, $\widetilde{M}^2$ is independent of $\xi$, and one recovers the CP conserving mass states.

Now we perform a first order perturbative expansion to find the following eigenvectors and eigenvalues of the matrix in Eq.\ref{mexd} in terms of the mass eigenstates of the CP conserving 2HDM\footnote{Note that this expansion works when $\lambda_{H,h}\neq {\lambda_6\over2}$, which excludes a small portion of the parameter space. One should also make sure that $(\lambda_6-\lambda_5)$ is not too large. This depends on $\tan\beta$ and Re($\mu_3^2$). In this paper, I assume $\tan\beta < 4$, which is shown to be preferred by a strongly first order phase transition \cite{Dorsch:2013wja}. This choice is also consistent with the perturbative expansion in the studied portion of the parameter space, namely the mass corrections are not larger than $10\%$. }:  
\begin{enumerate}
 \item \emph{Mostly CP-odd} eigenstate:
 \begin{subequations}\label{A}
  \begin{align}
   \widetilde{A}^0 &\simeq A^0+\frac{(\lambda_6-\lambda_5)s_\beta}{\lambda_6-2\lambda_H}\,\xi\,H^0+\frac{(\lambda_6-\lambda_5)c_\beta}{\lambda_6-2\lambda_h}\,\xi\,h^0 \\
   m^2_{\widetilde{A}}&\simeq\frac{\lambda_6}{2}v^2 + \frac{1}{2}(\lambda_6-\lambda_5)\left(-1+\frac{(\lambda_6-\lambda_5)s^2_\beta}{\lambda_6-2\lambda_H}+\frac{(\lambda_6-\lambda_5)c^2_\beta}{\lambda_6-2\lambda_h}\right)\xi^2\,v^2 \label{cpmA}
  \end{align}
 \end{subequations}
 \item Heavier \emph{mostly CP-even} eigenstate:
  \begin{subequations}
   \begin{align}
    \widetilde{H}^0 &\simeq H^0 + \frac{(\lambda_6-\lambda_5)s_\beta}{2\lambda_H-\lambda_6}\,\xi\,A^0  \\
    m_{\widetilde{H}}^2&\simeq\lambda_H\,v^2+\left(\epsilon_H + \frac{(\lambda_6-\lambda_5)^2s^2_\beta}{2(2\lambda_H-\lambda_6)}\right)\xi^2\,v^2
   \end{align}

  \end{subequations}
 \item Lighter \emph{mostly CP-even} eigenstate:
  \begin{subequations}
   \begin{align}
    \widetilde{h}^0 &\simeq h^0 + \frac{(\lambda_6-\lambda_5)c_\beta}{2\lambda_h-\lambda_6}\,\xi\,A^0  \\
    m_{\widetilde{h}}^2&\simeq\lambda_h\,v^2+\left(\epsilon_h + \frac{(\lambda_6-\lambda_5)^2c^2_\beta}{2(2\lambda_h-\lambda_6)}\right)\xi^2\,v^2
   \end{align}

  \end{subequations}
\end{enumerate}

\begin{figure}
 \includegraphics[scale=0.5]{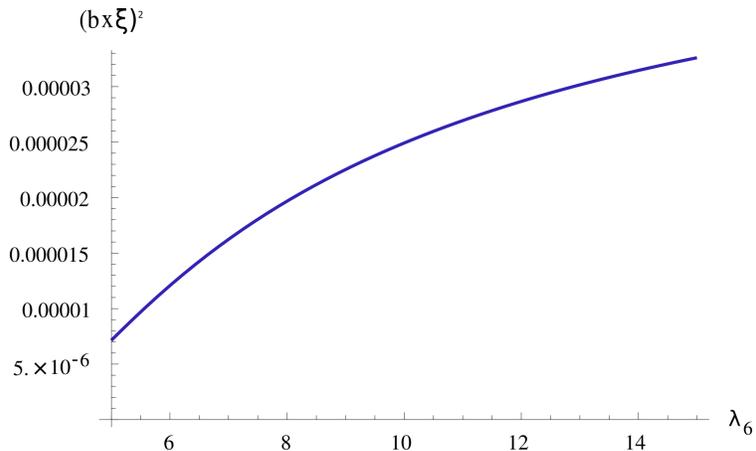}
 \caption{Probability of CP-odd decays of $\tilde{h}^0$ ($b=\frac{(\lambda_6-\lambda_5)c_\beta}{\lambda_6-2\lambda_h}$) vs $\lambda_6$ for $\tan\beta=1,\,\alpha-\beta=\frac{\pi}{2},\,m_{\tilde{H}}=500\,{\rm GeV},{\rm Re}(\mu_3^2)=5\times 10^{4}\,{\rm GeV}^2$} and $\xi=0.01$ \label{bxisl}
\end{figure}

 A few comments are in order at this point. In a generic CP violating 2HDM, there would be 3 angles corresponding to the mixing between the three neutral Higgs states. In the small CP expansion, we find that the angle that mixes $H^0$ and $h^0$ is zero. The other two mixing angles, those mix CP-even and CP-odd states, are both proportional to the small CP violating phase $\xi$. These results agree with Ref.\cite{ginzburg_symmetries_2004}. In the next section, we will constrain the CP violating phase $\xi$ from the electron EDM, and will find that $\xi \sim 0.01$ is easily allowed. Given that much CP-violation, I predict the observed 125 GeV Higgs particle $\tilde{h}^0$ would have $\mathcal{O}(0.001\%)$ CP-odd probability (see Fig.\ref{bxisl}). There are observables that would help to measure the CP mixing properties of Higgs \cite{godbole_aspects_2007}.

\section{CP violation and the finite temperature potential} \label{2hdmpt}

It is mentioned in the Introduction that EWBG requires a strongly first order phase transition. In order to study the behavior of the phase transition, one needs to use the finite temperature potential for the Higgs sector. Fortunately, the finite temperature potential has been calculated \cite{Dolan:1973qd} and used many times in the literature \cite{Fromme:2006cm, PhysRevD.47.3546,Davies:1994id,Dorsch:2013wja}. I will not go into any details of these calculations here, but I will comment on how to incorporate a small CP violation in the potential when searching for a first order phase transition. The following form of the scalar potential is the most useful for finding the finite temperature potential:
\begin{align}\label{2hpp}
 V_0 = -\mu_1^2&\Phi_1^\dagger\Phi_1-\mu_2^2\Phi_2^\dagger\Phi_2-\mu_3^2\Phi_1^\dagger\Phi_2-\mu_3^{2*}\Phi_2^\dagger\Phi_1 +h_1(\Phi_1^\dagger\Phi_1)^2+h_2(\Phi_2^\dagger\Phi_2)^2 \nonumber \\
 &+h_3(\Phi_1^\dagger\Phi_1)(\Phi_2^\dagger\Phi_2)+h_4|\Phi_1^\dagger\Phi_2|^2+\frac{h_5}{2}\left[(\Phi_1^\dagger\Phi_2)^2+(\Phi_2^\dagger\Phi_1)^2\right]
\end{align}
 Note that $\mu_3^2$ term softly breaks the aforementioned $Z_2$ symmetry. If $\mu_3^2$ is either real or purely imaginary, then there is no CP violation. The coupling constants $h_i$ ($i=1,\ldots,5$) are real and satisfy 
\begin{align}
 h_1>0,\hspace{.5in} h_2>0, \hspace{.5in} 2\sqrt{h_1\,h_2}+h_3>0,\hspace{.5in} 2\sqrt{h_1\,h_2}+h_3+h_4\pm h_5>0
\end{align}
so that the potential is bounded from below. These are the renormalized couplings at the EW scale. The renormalization group flow would drive these couplings to different values at different scales, and one should impose that the potential is at least metastable at the EW scale. This puts constraints on the coupling constants at the weak scale, and consequently bounds on the Higgs masses. Furthermore, the vacuum of 2HDMs is richer than the SM vacuum. For example one can have charge breaking or CP violating vacua. The stability of these minima are studied in Ref.\cite{Branco:2011iw} and references therein.

In order to write the coupling constants in terms of the (CP conserving) Higs masses, let us compare Eqn.\ref{2hp} and Eq.\ref{2hpp}:

\begin{subequations}\label{ccr}
 \begin{align}
  \mu_1^2&=\lambda_1\,v_1^2+\lambda_3\,v^2
	 ={1\over2}\left(m_h^2\,s_\alpha^2+m_H^2\,c_\alpha^2-2\,\re\tan\beta+\frac{m_H^2-m_h^2}{2}\,s_{2\alpha}\,\tan\beta \right)\\
  \mu_2^2& =\lambda_2\,v_2^2+\lambda_3\,v^2 
	 ={1\over2}\left(m_h^2\,c_\alpha^2+m_H^2\,s_\alpha^2-2\,\re\cot\beta+\frac{m_H^2-m_h^2}{2}\,s_{2\alpha}\,\cot\beta \right)\\
  {\rm Re}(\mu_3^2)&= {1\over2}\lambda_5\,v_1\,v_2\,\cos\xi \\
  {\rm Im}(\mu_3^2)&= -{1\over2}\lambda_6\,v_1\, v_2\,\sin\xi \\
  h_1&= \lambda_1+\lambda_3 
     =\frac{1}{4\,v^2\,c_\beta^2}\left[2m_h^2\,s_\alpha^2+2m_H^2\,c_\alpha^2-2\,\re\tan\beta\right]\\
  h_2&= \lambda_2+\lambda_3 
     =\frac{1}{4\,v^2\,s_\beta^2}\left[2m_h^2\,c_\alpha^2+2m_H^2\,s_\alpha^2-2\,\re\cot\beta\right]\\
  h_3&= 2\lambda_3+\lambda_4 
     ={1\over v^2}\left[2m_{H^\pm}^2+(m_H^2-m_h^2)\frac{s_{2\alpha}}{s_{2\beta}}-\frac{2\,\re}{s_{2\beta}}\right]\\
  h_4&= -\lambda_4+\frac{\lambda_5+\lambda_6}{2} 
     ={1\over v^2}\left[\frac{2\,\re}{s_{2\beta}}-2m_{H^\pm}^2+\frac{\lambda_6\,v^2}{2}\right] \\
  h_5&= \frac{\lambda_5-\lambda_6}{2}
     ={1\over v^2}\left[\frac{2\,\re}{s_{2\beta}}-\frac{\lambda_6\,v^2}{2}\right]
 \end{align}

\end{subequations}
We get two constraint equations by minimizing the potential:
\begin{align}
 -\mu_1^2\,v_1-{\rm Re}(e^{i\xi}\mu_3^2)\,v_2+h_1\,v_1^3+{h_3+h_4+h_5\cos(2\xi)\over2}\,v_2^2\,v_1=0 \\
 -\mu_2^2\,v_2-{\rm Re}(e^{i\xi}\mu_3^2)\,v_1+h_2\,v_2^3+{h_3+h_4+h_5\cos(2\xi)\over2}\,v_1^2\,v_2=0
\end{align}

I chose to keep $\lambda_6$ as one of the parameters instead of $m_A$ (remember that in the case of CP conservation, we have $m_A^2 = \frac{\lambda_6}{2}v^2$). A small amount of CP violation can be incorporated by the following replacements in Eq.\ref{ccr} : 
\begin{align}\label{Mm}
 m_h^2&\simeq m_{\tilde{h}}^2+\left( \frac{\lambda_6\,v^2}{4}-\frac{\,\re}{s_{2\beta}} \right)\cos2(\alpha+\beta)\,\xi^2 +\frac{\lambda_6\,v^2-4\,\re\,s_{2\beta}^{-1}}{2(\lambda_6\,v^2-2\,m_{\tilde{h}}^2)}\left[ m_{\tilde{h}}^2-\frac{2\,\re}{s_{2\beta}}+\frac{\lambda_6\,v^2}{2}c_{2\beta}-2\,\re\cot2\beta \right]\xi^2 \\
 m_H^2&\simeq m_{\tilde{H}}^2+\left( \frac{\lambda_6\,v^2}{4}-\frac{\,\re}{s_{2\beta}} \right)\cos2(\alpha+\beta)\,\xi^2 +\frac{\lambda_6\,v^2-4\,\re\,s_{2\beta}^{-1}}{2(\lambda_6\,v^2-2\,m_{\tilde{H}}^2)}\left[ m_{\tilde{H}}^2-\frac{2\,\re}{s_{2\beta}}-\frac{\lambda_6\,v^2}{2}c_{2\beta}+2\,\re\cot2\beta \right]\xi^2 
\end{align}
Recalling Eq. \ref{cpmA}, the mass of the \emph{mostly CP-odd} Higgs can be calculated from:
\begin{align} \label{mA}
 m_{\tilde{A}}^2\simeq\frac{\lambda_6\,v^2}{2}+\left(\frac{\lambda_6\,v^2}{2}-\frac{2\,\re}{s_{2\beta}}\right)\left[-1+\frac{s_\beta^2\,\lambda_6\,v^2-2\,\re\tan\beta}{\lambda_6\,v^2-2m_{\tilde{H}}^2}+\frac{c_\beta^2\,\lambda_6\,v^2-2\,\re\cot\beta}{\lambda_6\,v^2-2m_{\tilde{h}}^2}\right]\,\xi^2
\end{align}
Hence, the parameters used are $m_{\tilde{H}},m_{\tilde{h}},m_{H^\pm},\lambda_6,\beta,\alpha,\xi,$ and ${\rm Re}(\mu_3^2)$. In the next section, we will constrain the angle $\xi$ by using the electron EDM. This small CP violation is not expected to change the strength of the phase transition; however, it would affect the baryon number asymmetry in the universe.

\section{Electron EDM}\label{edmsec}
EDMs violate CP and T. Consequentially, they are a good handle on how much CP violation we can expect from new physics (NP) sources. In the SM, the only source of CP violation is in the quark sector via the CKM matrix\footnote{The strong CP problem that is caused by the $\theta G\tilde{G}$ term, which is a possible large CP violating term, is assumed to be solved by the axion.}. The small CP violating phase in the CKM matrix produces a small EDM for the neutron ($d_n\simeq 10^{-32}\,e\,{\rm cm}$) and an even smaller one for the electron ($d_e \leq 10^{-38}\,e\,{\rm cm}$) \cite{pospelov_electric_2005}. Experimental bounds, though very stringent, are orders of magnitude higher than SM predictions: $|d_n| < 3.3 \times 10^{-26}\,e\,{\rm cm}$ at $95\%$ confidence level (CL) \cite{jung_electric_2013} and $|d_e| < 0.87\times 10^{-28}\,e\,{\rm cm}$ at $90\%$ confidence level (CL) \cite{acme_collaboration_order_2013}. For good theoretical and experimental reviews on the EDMs, see \cite{pospelov_electric_2005, 
jung_electric_2013,fukuyama_searching_2012}.  Here I will only consider the electron EDM, 
because it gives more 
stringent constraints on the possible CP 
violation in 2HDMs \cite{shu_impact_2013}.
\begin{figure}[h]
 \includegraphics[scale=.8]{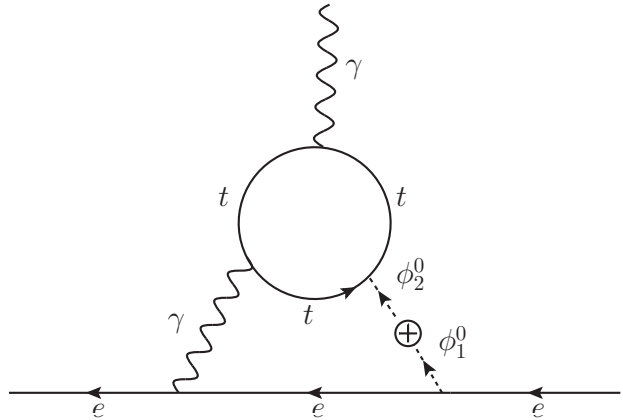}
 \caption{Top-loop diagram contributing to the electron EDM. Boson-loop diagrams are similarly two-loop diagrams but include $W/Z$ bosons in the loops (for example, exchanging the top-loop for a $W$-loop). } \label{edmt}
\end{figure}

The electron EDM in a 2HDM was calculated by Barr and Zee \cite{barr_electric_1990,barr_electric_1990_erratum}. The leading diagrams that contribute to the EDM, called the Barr-Zee diagrams, are two-loop diagrams involving either a top quark loop or a boson loop (see Figure.\ref{edmt}). In the models with a softly broken $Z_2$ symmetry, the CP violation only comes from the neutral Higgs bosons. Therefore, we do not need to include any diagrams without a neutral Higgs in the loops. The diagrams that include a $Z^0$ boson come  with a factor of ${1\over4}-\sin^2\theta_W$, which makes their contribution much smaller than the the diagrams that include a photon that couples to the electron. Hence, we will only consider the top-loop and $W$-loop  contributions to the EDM. These contributions are \cite{barr_electric_1990,barr_electric_1990_erratum}:
\begin{subequations}\label{edm}
\begin{align}
 \left[\frac{d_e}{e}\right]_t&=\left[\frac{16\alpha}{3(4\pi)^3}\sqrt{2}G_F m_e\right]\left\{ [f(x_{tH})+g(x_{tH})]{\rm Im}Z_0 - [f(x_{tH})-g(x_{tH})]{\rm Im}\tilde{Z_0}\right\} \label{edmtop}\\ 
 \left[\frac{d_e}{e}\right]_W &=-\left[\frac{4\alpha}{(4\pi)^3}\sqrt{2}G_F m_e\right]\bigl\{3 f(x_{WH})+5g(x_{WH})\bigr\}\sin^2\beta\,{\rm Im}Z_0 \label{edmW}
\end{align}
\end{subequations}
where $x_{iH}=\frac{m_i^2}{m_H^2}$ ($i=t,W$), $m_e$ is the electron mass, $\alpha$ is the fine structure constant (at scale $m_e$), $G_F$ is the Fermi constant, and 
\begin{align}
 f(z)&=\frac{1}{2}z \int_0^1 \frac{1-2x(1-x)}{x(1-x)-z}\ln(\frac{x(1-x)}{z}) \,dx \\
 g(z)&=\frac{1}{2}z \int_0^1 \frac{1}{x(1-x)-z}\ln(\frac{x(1-x)}{z})\, dx \\
 \frac{<\phi_2^0\phi_1^{0*}>}{v_1^* v_2} &\equiv\sum_n \sqrt{2}G_F Z_{0n} \frac{1}{q^2+m_{H_n}^2} \\
 \frac{<\phi_2^0\phi_1^{0}>}{v_1 v_2} &\equiv\sum_n \sqrt{2}G_F \tilde{Z}_{0n} \frac{1}{q^2+m_{H_n}^2}  
\end{align}
In the above equations, $n$ runs through the number of Higgs particles. If there is more than one Higgs boson, Eqn.\ref{edmtop} and \ref{edmW} become a sum over the different Higgs bosons. In the following analyses, we will not assume that the lightest Higgs dominates the loop contributions, so we will take all of the neutral Higgs bosons into account.

We can use the small CP-violation expansion from Section \ref{2hdmcpv} to write the correlators $<\phi_2^0\phi_1^{0*}>$ and $<\phi_2^0\phi_1^{0}>$ as follows: 
\begin{subequations}
 \begin{align}
  <\phi_2^0\phi_1^{0*}>\simeq &\frac{s_{2\beta}}{2}<\tilde{G}^0\tilde{G}^0> +\frac{1}{2}\left[-s_{2\beta}+2\xi^2(-2ab\,c_{2\alpha}+(a^2-b^2)s_{2\alpha})+2i\xi(a\cos(\alpha-\beta)+b\sin(\alpha-\beta))\right]<\tilde{A}^0\tilde{A}^0> \notag \\
  &+ \frac{1}{2} \left[ s_{2\alpha} - \xi^2\,a^2\,s_{2\beta} - 2i\xi\,a\,\cos(\alpha-\beta) \right] <\tilde{H}^0\tilde{H}^0> 
  + \frac{1}{2}\left[ -s_{2\alpha} - \xi^2\,b^2\,s_{2\beta} - 2i\xi\,b\,\sin(\alpha-\beta) \right] <\tilde{h}^0\tilde{h}^0> \\
  <\phi_2^0\phi_1^{0}>\simeq &-\frac{s_{2\beta}}{2}<\tilde{G}^0\tilde{G}^0> +\frac{1}{2}\left[s_{2\beta}+2\xi^2(-2ab\,c_{2\alpha}+(a^2-b^2)s_{2\alpha})+2i\xi(a\cos(\alpha+\beta)+b\sin(\alpha+\beta))\right]<\tilde{A}^0\tilde{A}^0> \notag \\
  &+ \frac{1}{2} \left[ s_{2\alpha} + \xi^2\,a^2\,s_{2\beta} - 2i\xi\,a\,\cos(\alpha+\beta) \right] <\tilde{H}^0\tilde{H}^0> + \frac{1}{2}\left[ -s_{2\alpha} + \xi^2\,b^2\,s_{2\beta} - 2i\xi\,b\,\sin(\alpha+\beta) \right] <\tilde{h}^0\tilde{h}^0>
 \end{align}
\end{subequations}
where $a=\frac{(\lambda_6-\lambda_5)s_\beta}{\lambda_6-2\lambda_H}$ and $b=\frac{(\lambda_6-\lambda_5)c_\beta}{\lambda_6-2\lambda_h}$. Note that the Nambu-Goldstone boson terms in the correlators are real, and hence they do not contribute to the EDMs. Then Eqn.\ref{edm} becomes:
\begin{subequations}\label{edm2}
\begin{align}
 \left[\frac{d_e}{e}\right]_t&=\xi\left(\frac{32\alpha}{3(4\pi)^3}\sqrt{2}G_F\,m_e\right)\biggl\{ s_\alpha\left(a\,\frac{f(x_{tA})-f(x_{tH})}{\cos\beta} + b\,\frac{g(x_{tA})-g(x_{th})}{\sin\beta}\right)+c_\alpha\left(a\,\frac{g(x_{tA})-g(x_{tH})}{\sin\beta} - \,b\frac{f(x_{tA})-f(x_{th})}{\cos\beta}\right) \biggr\} \\
 \left[\frac{d_e}{e}\right]_W&= -\xi\tan\beta\left(\frac{4\alpha}{(4\pi)^3}\sqrt{2}G_F\,m_e\right)\biggl\{\bigl[ 3f(x_{W\tilde{A}}) + 5g(x_{W\tilde{A}})\bigr]\bigl[ a\cos(\alpha-\beta)+b\sin(\alpha-\beta)\bigr]\biggr. \notag \\
\biggl. & \hspace{1.9 in} - a \bigl[ 3f(x_{W\tilde{H}}) + 5g(x_{W\tilde{H}})\bigr]\cos(\alpha-\beta)
 - b\bigl[ 3f(x_{W\tilde{h}}) + 5g(x_{W\tilde{h}})\bigr]\sin(\alpha-\beta) \biggr\}
\end{align}
\end{subequations}

\begin{figure}
 \includegraphics[scale=.5]{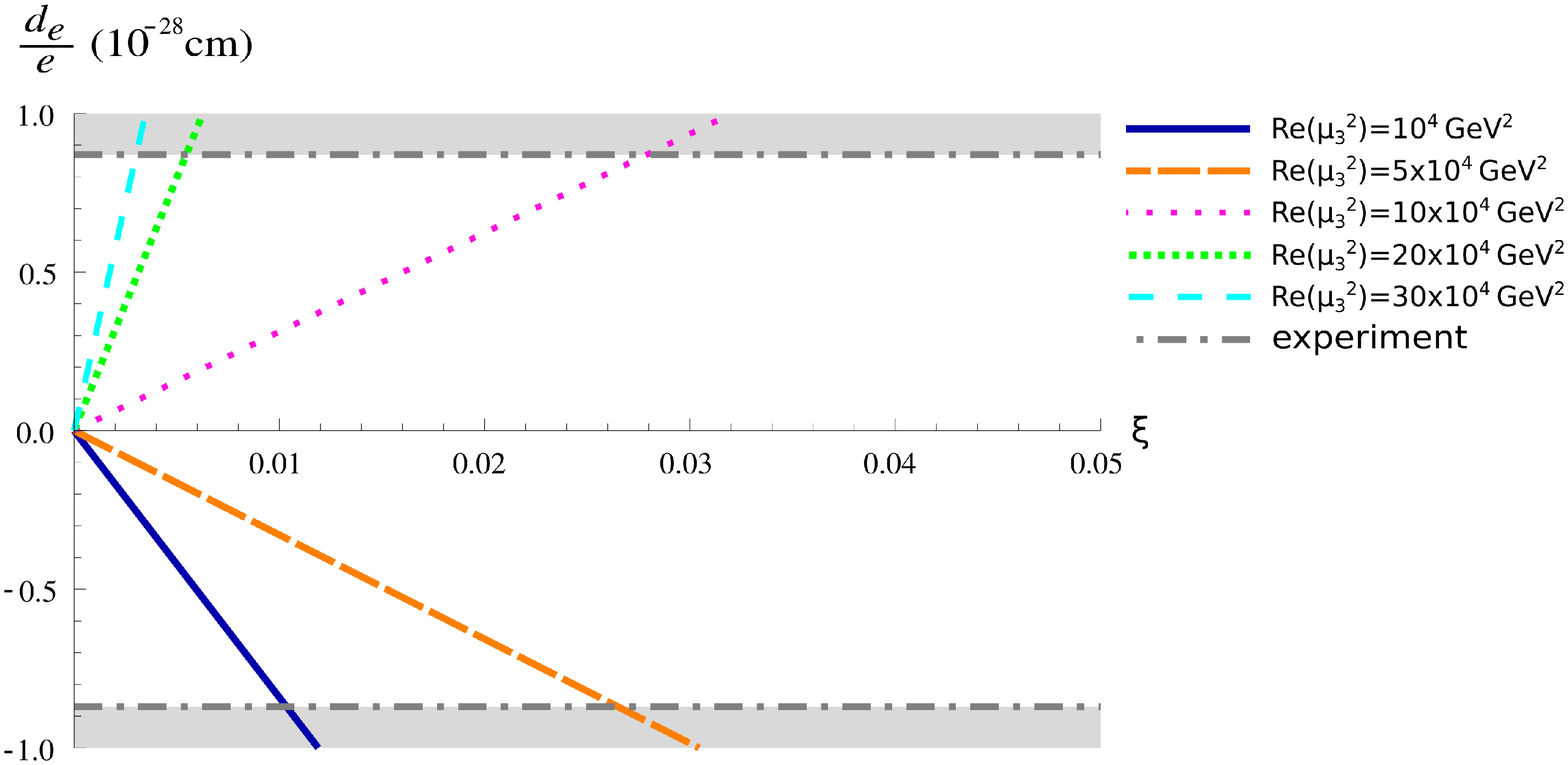}
 \caption{Electron EDM vs $\xi$ plot for $\lambda_6=5$ ($m_{\tilde{A}} \simeq 375-405$ GeV), $m_{\tilde{H}} =500$ GeV, , $\tan\beta=1$, $\alpha-\beta={\pi\over2}$. The gray region is experimentally excluded.} \label{l5mu}
\end{figure}

\begin{figure}
 \includegraphics[scale=.5]{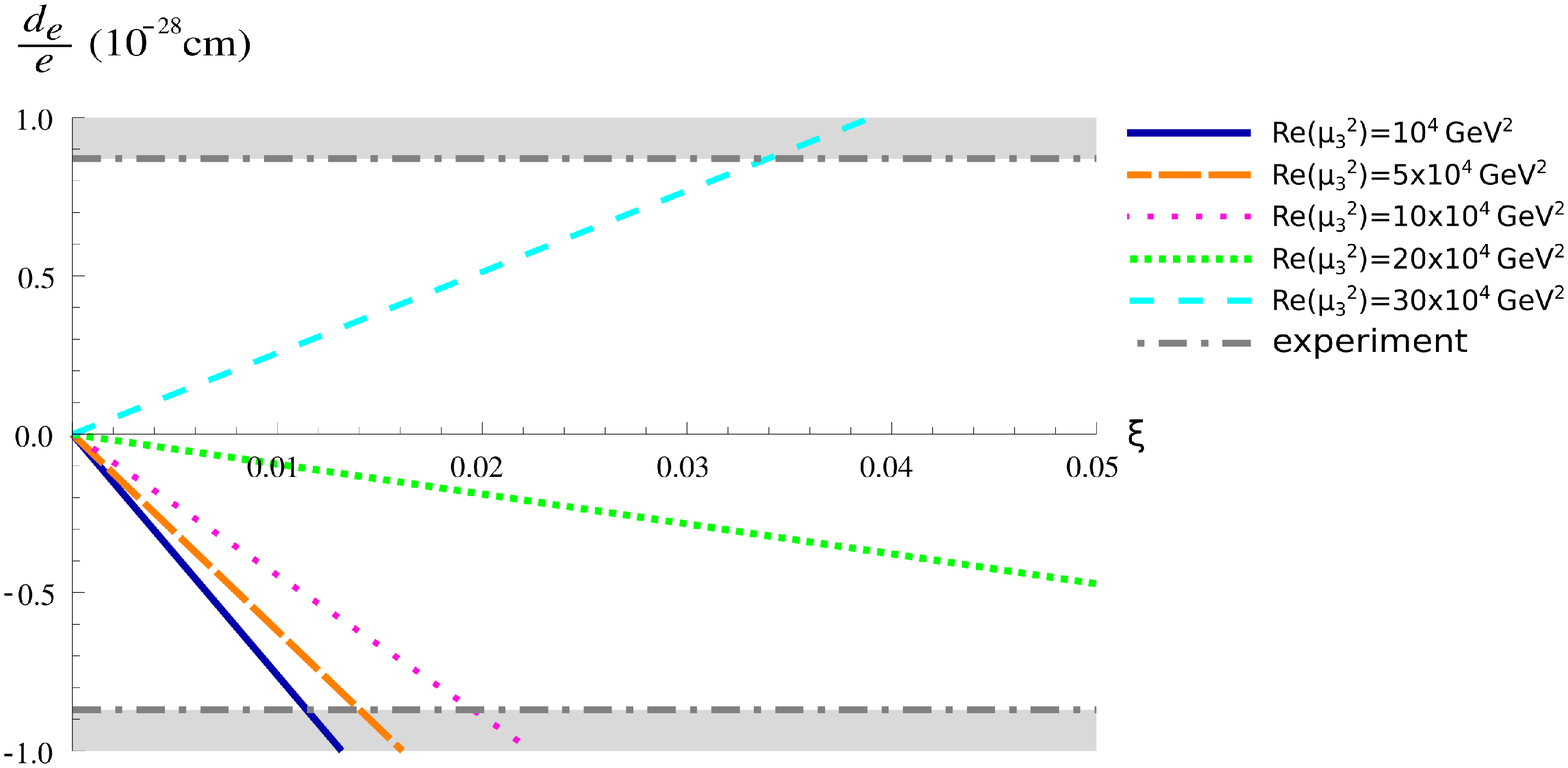}
 \caption{Electron EDM vs $\xi$ plot for $\lambda_6=15$ ($m_{\tilde{A}} \simeq 670-690$ GeV), $m_{\tilde{H}} =500$ GeV, , $\tan\beta=1$, $\alpha-\beta={\pi\over2}$. The gray region is experimentally excluded.} \label{l15mu}
\end{figure}

\begin{figure}[h]
\centering
\subfloat[$\lambda_6=5, \, {\rm Re} (\mu_3^2)=10^4\,{\rm GeV}^2$]{
\includegraphics[scale=0.43]{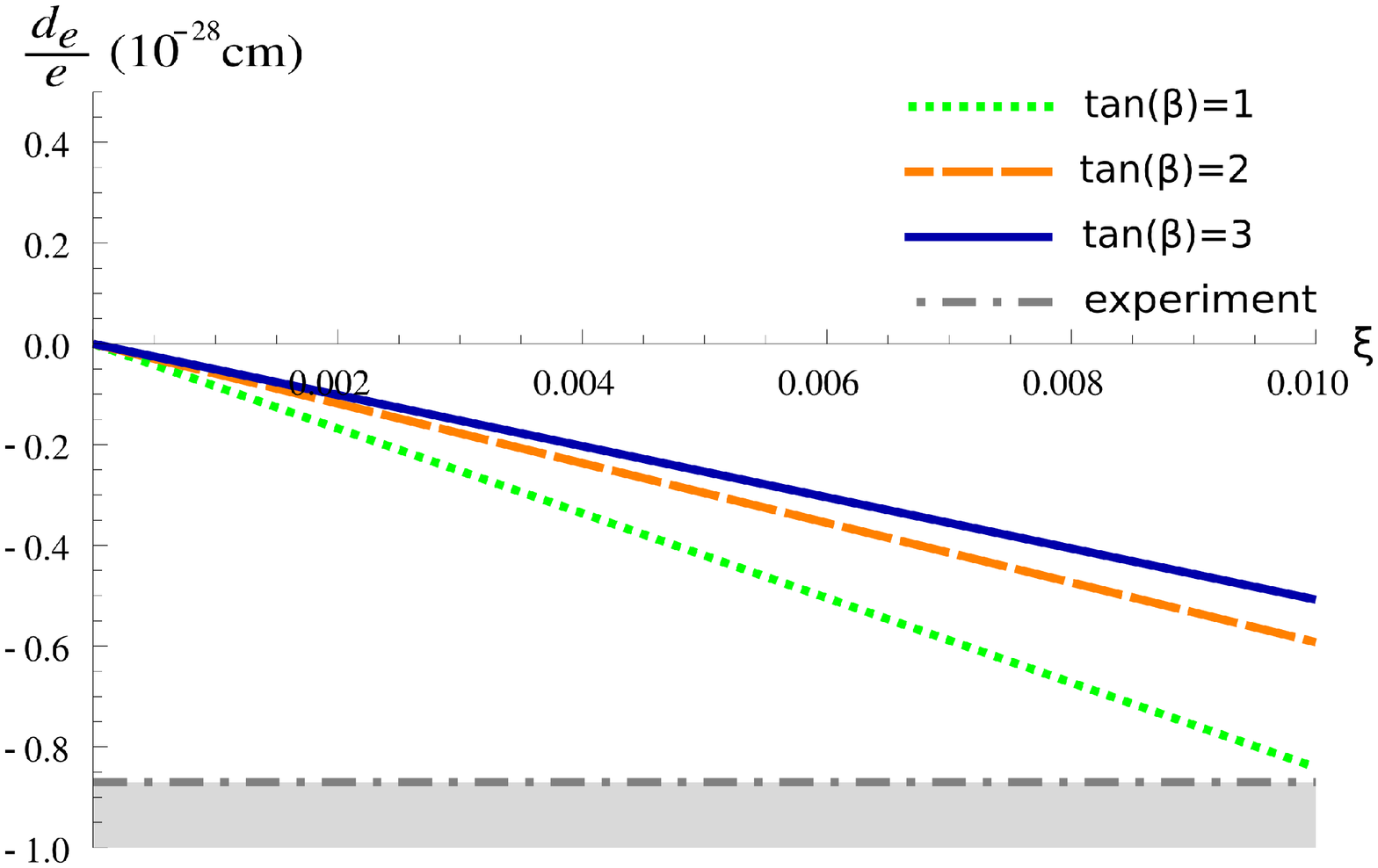}
\label{l5beta}
}
\subfloat[$\lambda_6=15, \, {\rm Re} (\mu_3^2)=5\times10^4\,{\rm GeV}^2$]{
\includegraphics[scale=0.43]{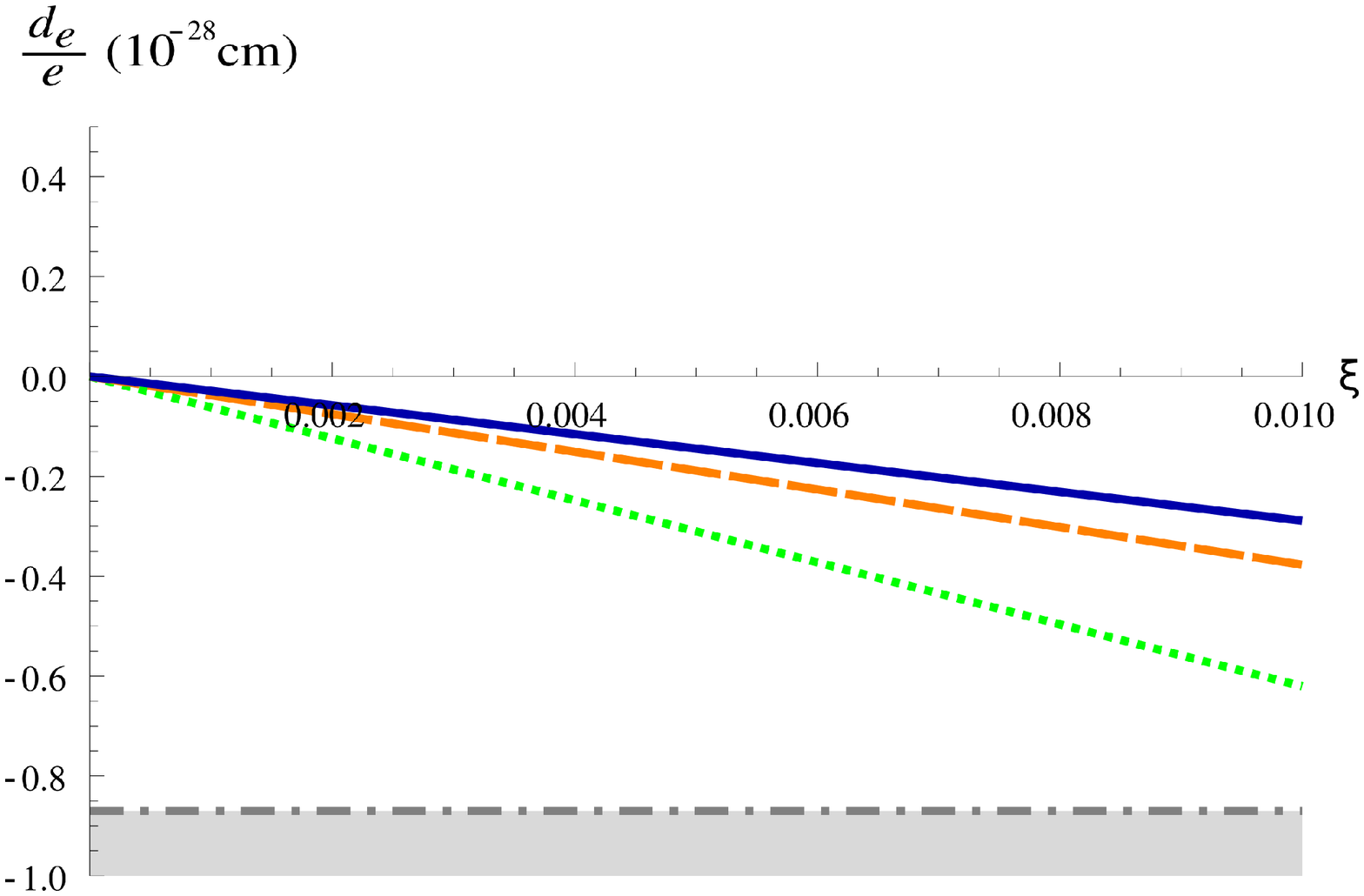}
 \label{l15beta}
}
\caption{Electron EDM vs $\xi$ plots for $\tan\beta = 1-3$ and for (a) $\lambda_6=5, \, {\rm Re} (\mu_3^2)=10^4\,{\rm GeV}^2$ and (b) $\lambda_6=15, \, {\rm Re} (\mu_3^2)=5\times10^4\,{\rm GeV}^2$. The following parameters are used in both plots: $m_{\tilde{H}}=500$ GeV, $\alpha-\beta={\pi\over2}$ }\label{lbeta}
\end{figure}

This is a good point to remind ourselves of the parameters we are using in this model: the charged Higgs mass, $m_{H^+}$, the heavier mostly CP-even neutral Higgs mass, $m_{\tilde{H}}$, the lighter mostly CP-even neutral Higgs mass, $m_{\tilde{h}} (= 125{\rm GeV})$, $\tan\beta (\equiv \frac{v_2}{v_1})$, $\alpha (=\beta + {\pi\over2})$, the coupling constant $\lambda_6$ (which is closely related to the mass of the mostly CP-odd Higgs mass), and ${\rm Re}(\mu_3^2)$. We can see that the top- and the $W$-loop contributions to the electron EDM do not depend on the charged Higgs mass. This is expected since the CP violation is mediated by the neutral scalar sector in this model. I chose the charged Higgs mass to be degenerate with $\tilde{H}^0$, since this choice also satisfies electroweak precision constraints \cite{Dorsch:2013wja}.\footnote{Actually, it is enough to have a charged Higgs degenerate with any of the neutral scalars to satisfy the EW precision constraints. However, there are other 
constraints on the charged Higgs mass, for example coming from $B$-meson decays, which puts a lower bound of $\sim300$ GeV on the allowed mass \cite{hermann_barbx_s_2012}. Hence, the charged Higgs can not be degenerate with the 125 GeV Higgs.} The dependence of the electron EDM on the rest of the parameters can be seen in Fig.\ref{l5mu}-\ref{lbeta}. There is cancellation between the top- and the $W$-loop in some portions of the parameter space, and an $\mathcal{O}(10^{-2})$  CP violation is allowed. Some general observations are:
\begin{itemize}
 \item It can be seen from Eqn.\ref{edm2} that the electron EDM is proportional to $\lambda_6-\lambda_5\equiv \lambda_6-\frac{2{\rm Re}(\mu_3^2)}{v^2s_\beta c_\beta}$. In order to have a small EDM, there should be good cancellation between $\lambda_6 v^2$ and ${\rm Re}(\mu_3^2)$. Hence, larger ${\rm Re}(\mu_3^2)$ would require larger $\lambda_6$ (heavier $m_{\tilde{A}}$). This can be seen in Fig.\ref{l5mu} and \ref{l15mu}.  For example, for ${\rm Re}(\mu_3^2)=20\times10^4\,{\rm GeV}^2$ (which gives $\mu_3 \simeq 450$ GeV), $\xi=0.1$ with an $\tilde{A}^0$ of mass $\sim 680$ GeV is allowed. These parameters also give a strongly first order EW phase transition \cite{Dorsch:2013wja}. 
 
 \item For smaller values of $\mu_3$ ($< 250$ GeV), lighter $\tilde{A}^0$ ($m_{\tilde{A}}\simeq 400$ GeV) allows larger $\xi$. Ref \cite{Dorsch:2013wja} shows that $\mu_3 \simeq250$ GeV is allowed, while $m_{\tilde{A}}<400$ GeV hardly gives a first order phase transition. Including the CP violation could eliminate more of the parameter space that allows a strongly first order phase transition. 

 \item The effect of $\tan\beta$ on the EDM depends on the parameters, \emph{e.g.} $\alpha$, and the sign of $\lambda_6-\frac{2{\rm Re}(\mu_3^2)}{v^2s_\beta c_\beta}$ (see Fig.\ref{lbeta}). Furthermore, a strong deviation from or preference of $\alpha-\beta={\pi\over2}$ is not found throughout the examined parameter space (figures not shown).

 \item If $\alpha-\beta=\frac{\pi}{2}$ (and $m_{\tilde{H}} \gg m_t$), then the electron EDM does not depend on $m_{\tilde{H}}$. When $\alpha-\beta \neq \frac{\pi}{2}$, we find that as one raises $m_{\tilde{H}}$ beyond 800 GeV, the dependence of EDM on $m_{\tilde{H}}$ becomes negligible (figures not shown).
\end{itemize}

We see that $\xi=0.01-0.05$ (up to $\xi\simeq0.1$ in some cases) is allowed in a 2HDM by the electron EDM bounds. \footnote{Note that this phase would not affect the CP violation in $B-$ and $K-$ meson systems. The CKM phase is of $\mathcal{O}(1)$, and is still the main source of CP violation in neutral meson systems. The CKM phase is not enough to explain the baryon asymmetry because the CP violating processes involve all three generations of quarks, and hence are suppressed by small Yukawa couplings.} Remember that we are introducing CP violation in the Higgs sector to achieve the baryon asymmetry in the universe. The relation between the baryon asymmetry and the CP violating phase is known only approximately \cite{Cohen:1993nk} and definitely needs more cautious studies\footnote{For good reviews, see \cite{morrissey_electroweak_2012, konstandin_quantum_2013} and references therein}. The estimate of Ref.\cite{Cohen:1993nk} gives $\xi \sim 10^{-2}$. This estimate suggests that we are very close to seeing a 
non-zero electron EDM.

\section{Conclusion} \label{conc}

The baryon asymmetry of the universe cannot be explained by the SM EWBG with a Higgs boson of mass 125 GeV, because the EW transition is not a first order phase transition. Furthermore, there is not enough CP violation in the SM. 2HDMs are interesting extensions of the SM, since they allow a strongly first order phase transition with a 125 GeV Higgs boson. Another feature of 2HDMs is that the scalar sector can have CP violation. This CP violation significantly complicates analyses by mixing CP-even and CP-odd Higgs fields. Due to EDM constraints, one expects only small CP violation. I assumed a small CP violating phase, $\xi$, in the Higgs sector, and expanded the Higgs mass states in terms of this small phase. This perturbative expansion makes it easier to incorporate CP violation in the studies that scan the 2HDM parameter space to look for a strongly first order phase transition. A small CP violation is not expected to affect the strength of the parameter transition. However, it can help 
constrain the phase space, since some portions of the parameter space that give a strongly first order parameter space would also give rise to large EDMs.   

I also used the analytic results for the Higgs states to constrain the phase $\xi$ using electron EDM data. With the current experimental bounds ($|d_e|<0.87\times10^{-28}\, e\,{\rm cm}$) we can accommodate $\xi\simeq0.01$ in most of the parameter space studied. Some studies suggest that one only needs $\xi \sim \mathcal{O}(10^{-2})$ to get sufficient baryon asymmetry. If a factor of two increase in the EDM sensitivity does not see a non-zero result, 2HDMs become fine-tuned. In addition, we need better estimates of the relation between the baryon asymmetry and the CP violating phase.

\begin{acknowledgments}
I would like to thank Ann Nelson for inspiring this work and for our many fruitful conversations, and David McKeen for his helpful comments on the manuscript. This work was supported in part by the U.S. Department of Energy under Grant No. DE-FG02-96ER40956.
\end{acknowledgments}

\appendix
\section{Perturbative calculations}\label{pert}
Let us start with the mass matrix from Eqn.\ref{matrix}:
\begin{align}
 \mathcal{M}^2=v^2\left(\begin{array}{cccc}
                             \frac{1}{2}(\lambda_5s_\xi^2+\lambda_6c_\xi^2)s_\beta^2 & -\frac{\lambda_6}{2}s_\beta c_\beta c_\xi & \frac{1}{2}(\lambda_5-\lambda_6)s_\beta^2 s_\xi c_\xi & \frac{\lambda_5}{2}s_\beta c_\beta s_\xi \\
                             -\frac{\lambda_6}{2}s_\beta c_\beta c_\xi & 2(\lambda_2+\lambda_3)s_\beta^2 s_\xi^2+\frac{\lambda_6}{2}c_\beta^2 & \left(2\lambda_3+\frac{\lambda_6}{2}\right)s_\beta c_\beta s_\xi & 2(\lambda_2+\lambda_3)s_\beta^2s_\xi c_\xi  \\
                             \frac{1}{2}(\lambda_5-\lambda_6)s_\beta^2s_\xi c_\xi & \left(2\lambda_3+\frac{\lambda_6}{2}\right)s_\beta c_\beta s_\xi & 2(\lambda_1+\lambda_3)c_\beta^2 + \frac{1}{2}(\lambda_5c_\xi^2+\lambda_6s_\xi^2)s_\beta^2 & \left(2\lambda_3+\frac{\lambda_5}{2}\right)s_\beta c_\beta c_\xi \\
                             \frac{\lambda_5}{2}s_\beta c_\beta s_\xi & 2(\lambda_2+\lambda_3)s_\beta^2s_\xi c_\xi & \left(2\lambda_3+\frac{\lambda_5}{2}\right)s_\beta c_\beta c_\xi & 2(\lambda_2+\lambda_3)s_\beta^2 c_\xi^2 + \frac{\lambda_5}{2}c_\beta^2
                            \end{array}\right)
\end{align}
in the basis $({\rm Im}\poz,\,\,\,{\rm Im}\ptz,\,\,\,{\rm Re}\poz,\,\,\,{\rm Re}\ptz)$. This matrix has a zero eigenvalue corresponding to the eigenvector:
\begin{align}
 \tilde{G}^0=\cos\beta\, {\rm Im}\poz + \sin\beta\phi_2^{0'}
\end{align}
where $\phi_2^{0'}=e^{-i\xi}\ptz$ (I will drop the primes from now on). We can get rid of the part of the mass matrix that corresponds to this zero eigenvalue by rotating it with the following rotation matrix:
\begin{align}
 R=\left(\begin{array}{cccc}
          \cos\beta & \sin\beta\,\cos\xi & 0 & -\sin\beta\,\sin\xi \\
          -\sin\beta & \cos\beta\,\cos\xi & 0 & -\cos\beta\,\sin\xi \\
          0 & 0 & 1 & 0 \\
          0 & \sin\xi & 0 & \cos\xi
         \end{array}\right)
\end{align}
The basis vector also rotates to become:
\begin{align}
 \left(\begin{array}{c}
{\rm Im}\poz\\
{\rm Im}\ptz\\
{\rm Re}\poz\\
{\rm Re}\ptz
\end{array}\right) \rightarrow \left(\begin{array}{c}
\tilde{G}^0\\
A^0\\
{\rm Re}\poz\\
{\rm Re}\phi_2^{0}
\end{array}\right)
\end{align}
where $A^0=-\sin\beta\,{\rm Im}\poz+\cos\beta\,{\rm Im}\phi_2^{0}$. After the rotation, the mass matrix $\mathcal{M}$ reduces to a $3\times3$ matrix in the basis $(A^0, \,\,\,{\rm Re}\poz,\,\,\,{\rm Re}\ptz )$:
\begin{align}\label{massm}
 M_\xi^2=v^2\left(\begin{array}{ccc}
                   {1\over4}(\lambda_5+\lambda_6+(\lambda_6-\lambda_5)c_{2\xi}) & \frac{1}{2}(\lambda_6-\lambda_5)s_\beta\,c_\xi\,s_\xi & \frac{1}{2}(\lambda_6-\lambda_5)c_\beta\,c_\xi\,s_\xi \\
                   \frac{1}{2}(\lambda_6-\lambda_5)s_\beta\,c_\xi\,s_\xi & 2(\lambda_1+\lambda_3)c_\beta^2+\frac{1}{2}(\lambda_5c_\xi^2+\lambda_6s_\xi^2)s_\beta^2 & \left[\left(2\lambda_3+\frac{\lambda_5}{2}\right)c_\xi^2+\left(2\lambda_3+\frac{\lambda_6}{2}\right)s_\xi^2\right]s_\beta\,c_\beta \\
                   \frac{1}{2}(\lambda_6-\lambda_5)c_\beta\,c_\xi\,s_\xi & \left[\left(2\lambda_3+\frac{\lambda_5}{2}\right)c_\xi^2+\left(2\lambda_3+\frac{\lambda_6}{2}\right)s_\xi^2\right]s_\beta\,c_\beta & 2(\lambda_2+\lambda_3)s_\beta^2+\frac{1}{2}(\lambda_5c_\xi^2+\lambda_6s_\xi^2)c_\beta^2
                  \end{array} \right)
\end{align}
 We will find approximate solutions for the eigenvectors and the eigenvalues of this matrix. In order to do that, let us assume CP violation is small, and expand $M_\xi$ in terms of $\xi$:
\begin{align}\label{mex}
  \widetilde{M}^2=v^2\left(\begin{array}{ccc}
                   \frac{\lambda_6}{2}\textcolor{red}{-\frac{1}{2}(\lambda_6-\lambda_5)\xi^2} & \textcolor{blue}{\frac{1}{2}(\lambda_6-\lambda_5)s_\beta\,\xi} & \textcolor{blue}{\frac{1}{2}(\lambda_6-\lambda_5)c_\beta\,\xi} \\
                   \textcolor{blue}{\frac{1}{2}(\lambda_6-\lambda_5)s_\beta\,\xi} & 2(\lambda_1+\lambda_3)c_\beta^2+\frac{\lambda_5}{2}\,s_\beta^2\textcolor{red}{+\frac{1}{2}(\lambda_6-\lambda_5)\,s_\beta^2\xi^2} & \left(2\lambda_3+\frac{\lambda_5}{2}\right)\,s_\beta\,c_\beta\textcolor{red}{+\frac{1}{2}(\lambda_6-\lambda_5)\,s_\beta\,c_\beta\,\xi^2} \\
                   \textcolor{blue}{\frac{1}{2}(\lambda_6-\lambda_5)c_\beta\,\xi} & \left(2\lambda_3+\frac{\lambda_5}{2}\right)\,s_\beta\,c_\beta\textcolor{red}{+\frac{1}{2}(\lambda_6-\lambda_5)\,s_\beta\,c_\beta\,\xi^2} & 2(\lambda_2+\lambda_3)s_\beta^2+\frac{\lambda_5}{2}\,c_\beta^2\textcolor{red}{+\frac{1}{2}(\lambda_6-\lambda_5)\,c_\beta^2\xi^2}
                  \end{array} \right)
\end{align}
The $2\times2$ bottom-right corner of this matrix is the same as $M^2$ (in Eq.\ref{mnocp}) plus an $\mathcal{O}(\xi^2)$ part. And one can see that its eigenvalues are the eigenvalues of $M^2$ with an $\mathcal{O}(\xi^2)$ correction. I will denote these eigenvalues as $v^2(\lambda_H+\epsilon_H\xi^2)$ and $v^2(\lambda_h+\epsilon_h\xi^2)$ with the definitions:
\begin{align}
 \lambda_H&={1\over4}\left(4\lambda_3+\lambda_5+4(\lambda_2\,s^2_\beta+\lambda_1\,c^2_\beta)+\sqrt{(4\lambda_3+\lambda_5)s^2_{2\beta}+[(4\lambda_3-\lambda_5)c_{2\beta}-4(\lambda_2\,s^2_\beta-\lambda_1\,c^2_\beta)]^2}\right) \\
 \epsilon_H&= \frac{1}{2}(\lambda_6-\lambda_5)\sin^2(\alpha+\beta)
\end{align}
and
\begin{align}
  \lambda_h&={1\over4}\left(4\lambda_3+\lambda_5+4(\lambda_2\,s^2_\beta+\lambda_1\,c^2_\beta)-\sqrt{(4\lambda_3+\lambda_5)s^2_{2\beta}+[(4\lambda_3-\lambda_5)c_{2\beta}-4(\lambda_2\,s^2_\beta-\lambda_1\,c^2_\beta)]^2}\right) \\
 \epsilon_h&=\frac{1}{2}(\lambda_6-\lambda_5)\cos^2(\alpha+\beta)
\end{align}
The angle $\alpha$ is same as before up to an $\mathcal{O}(\xi^2)$ correction which we do not keep:
\begin{align}
 \sin2\alpha=\frac{(4\lambda_3+\lambda_5)\sin2\beta}{\lambda_H-\lambda_h} + \mathcal{O}(\xi^2)
\end{align} 
Hence, the matrix in Eqn.\ref{mex} can be reduced to:
\begin{align}\label{mexdapp}
  \widetilde{M}=v^2\left(\begin{array}{ccc}
                   \frac{\lambda_6}{2}\textcolor{red}{-\frac{1}{2}(\lambda_6-\lambda_5)\xi^2} & \textcolor{blue}{\frac{1}{2}(\lambda_6-\lambda_5)s_\beta\,\xi} & \textcolor{blue}{\frac{1}{2}(\lambda_6-\lambda_5)c_\beta\,\xi} \\
                   \textcolor{blue}{\frac{1}{2}(\lambda_6-\lambda_5)s_\beta\,\xi} & \lambda_H\textcolor{red}{+\epsilon_H\xi^2} & 0 \\
                   \textcolor{blue}{\frac{1}{2}(\lambda_6-\lambda_5)c_\beta\,\xi} & 0 & \lambda_h\textcolor{red}{+\epsilon_h\xi^2}
                  \end{array} \right)
\end{align}
in the approximate basis $(A^0,\,\, H^0,\,\, h^0)$. Now, I will make some ans\"{a}tze to find the eigenvalues and eigenvectors of this matrix.
\begin{enumerate}
 \item My first ansatz is for the \emph{mostly CP-odd} Higgs: $\widetilde{A}^0 = (1,\,\, b_1\xi,\,\, c_1\xi)$ with eigenvalue $\left(\frac{\lambda_6}{2}+\alpha_1\xi^2\right)v^2$. Remember that $\frac{\lambda_6}{2}\,v^2$ is the mass of the CP-odd Higgs if there is no CP-violation in the potential. We can solve for $b_1,c_1$ and $\alpha_1$ from the eigenvalue equation:
 \begin{align}
  v^2\left(\begin{array}{ccc}
                   \frac{\lambda_6}{2}-\frac{1}{2}(\lambda_6-\lambda_5)\xi^2 &\frac{1}{2}(\lambda_6-\lambda_5)s_\beta\,\xi &\frac{1}{2} (\lambda_6-\lambda_5)c_\beta\,\xi \\
                   \frac{1}{2}(\lambda_6-\lambda_5)s_\beta\,\xi & \lambda_H+\epsilon_H\xi^2 & 0 \\
                  \frac{1}{2}(\lambda_6-\lambda_5)c_\beta\,\xi & 0 & \lambda_h+\epsilon_h\xi^2
                  \end{array} \right)\left(\begin{array}{l}
					     1\\
					     b_1\xi \\
					     c_1\xi
					     \end{array}\right)\simeq\left(\frac{\lambda_6}{2}+\alpha_1\xi^2\right)v^2\left(\begin{array}{l}
												  1\\
												  b_1\xi \\
												  c_1\xi
												  \end{array}\right)
 \end{align}
 Then we have 3 equations for 3 unknowns:
 \begin{align}
  \frac{\lambda_6}{2}-\frac{1}{2}(\lambda_6-\lambda_5)\xi^2+&\frac{1}{2}(\lambda_6-\lambda_5)s_\beta\xi^2\,b_1+\frac{1}{2}(\lambda_6-\lambda_5)c_\beta\xi^1\,c_1\simeq\frac{\lambda_6}{2}+\alpha_1\xi^2 \nonumber\\
  &\alpha_1\simeq \frac{1}{2}(\lambda_6-\lambda_5)(-1+s_\beta\,b_1+c_\beta\,c_1) \\
  \frac{1}{2}(\lambda_6-\lambda_5)s_\beta\xi+&(\lambda_H+\epsilon_H\xi^2)\xi\,b_1\simeq \left(\frac{\lambda_6}{2}+\alpha_1\xi^2\right)\xi\,b_1 \nonumber\\
  &b_1\simeq\frac{(\lambda_6-\lambda_5)s_\beta}{\lambda_6-2\lambda_H}\\
  \frac{1}{2}(\lambda_6-\lambda_5)c_\beta\xi+&(\lambda_h+\epsilon_h\xi^2)\xi\,c_1\simeq \left(\frac{\lambda_6}{2}+\alpha_1\xi^2\right)\xi\,c_1 \nonumber\\
  &c_1\simeq\frac{(\lambda_6-\lambda_5)c_\beta}{\lambda_6-2\lambda_h}
 \end{align}
So the \emph{mostly CP-odd} eigenstate is:
\begin{align}
 \widetilde{A}^0 \simeq A^0+\frac{(\lambda_6-\lambda_5)s_\beta}{\lambda_6-2\lambda_H}\xi\,H^0+\frac{(\lambda_6-\lambda_5)c_\beta}{\lambda_6-2\lambda_h}\xi\,h^0
\end{align}
with mass
\begin{align}
 m^2_{\widetilde{A}^0}\simeq\frac{\lambda_6}{2}\,v^2 + \frac{1}{2}(\lambda_6-\lambda_5)\left(-1+\frac{(\lambda_6-\lambda_5)s^2_\beta}{\lambda_6-2\lambda_H}+\frac{(\lambda_6-\lambda_5)c^2_\beta}{\lambda_6-2\lambda_h}\right)\xi^2\,v^2
\end{align}

\item My second ansatz is the heavy \emph{mostly CP-even} Higgs: $\widetilde{H}^0=(a_2\xi,\,\,1,\,\,c_2\xi^2)$ with eigenvalue $(\lambda_H+\alpha_2\xi^2)v^2$. The eigenvalue equation is:
 \begin{align}
   v^2\left(\begin{array}{ccc}
                   \frac{\lambda_6}{2}-\frac{1}{2}(\lambda_6-\lambda_5)\xi^2 &\frac{1}{2}(\lambda_6-\lambda_5)s_\beta\,\xi &\frac{1}{2} (\lambda_6-\lambda_5)c_\beta\,\xi \\
                   \frac{1}{2}(\lambda_6-\lambda_5)s_\beta\,\xi & \lambda_H+\epsilon_H\xi^2 & 0 \\
                  \frac{1}{2}(\lambda_6-\lambda_5)c_\beta\,\xi & 0 & \lambda_h+\epsilon_h\xi^2
                  \end{array} \right)\left(\begin{array}{l}
					     a_2\xi\\
					     1 \\
					     c_2\xi^2
					     \end{array}\right)\simeq(\lambda_H+\alpha_2\xi^2)v^2\left(\begin{array}{l}
												  a_2\xi\\
												  1 \\
												  c_2\xi^2
												  \end{array}\right)
 \end{align}
 This gives the following equations to solve for $a_2,c_2$ and $\alpha_2$:
 \begin{align}
  \frac{\lambda_6}{2}\xi\,a_2 +& \frac{1}{2}(\lambda_6-\lambda_5)s_\beta\xi\simeq\lambda_H\xi\,a_2 \nonumber \\
  &a_2\simeq\frac{(\lambda_6-\lambda_5)s_\beta}{2\lambda_H-\lambda_6} \\
  \frac{1}{2}(\lambda_6-\lambda_5)&s_\beta\xi^2\,a_2 +\lambda_H+\epsilon_H\xi^2=\lambda_H+\alpha_2\xi^2 \nonumber \\
  &\alpha_2\simeq\epsilon_H+\frac{1}{2}\frac{(\lambda_6-\lambda_5)^2s^2_\beta}{2\lambda_H-\lambda_6} \\
  \frac{1}{2}(\lambda_6-\lambda_5)&c_\beta\xi^2\,a_2+\lambda_h\,c_2\,\xi^2\simeq \lambda_H\xi^2\, c_2 \nonumber \\
  &c_2\simeq \frac{(\lambda_6-\lambda_5)^2c_\beta s_\beta}{2(2\lambda_H-\lambda_6)(\lambda_H-\lambda_{h})}
 \end{align}
So the heavier \emph{mostly CP-even} state is:
\begin{align}
 \widetilde{H}^0 \simeq H^0 + \frac{(\lambda_6-\lambda_5)s_\beta}{2\lambda_H-\lambda_6}\xi\,A^0 
\end{align}
with mass
\begin{align}
 m_{\widetilde{H}}^2\simeq\lambda_H\,v^2+\epsilon_H\xi^2\,v^2 + \frac{(\lambda_6-\lambda_5)^2s^2_\beta}{2(2\lambda_H-\lambda_6)}\xi^2\,v^2
\end{align}

\item My third ansatz is the lighter \emph{mostly CP-even} Higgs: $\tilde{h}^0=(a_3\xi,\,\,b_3\xi^2,\,\,1)$ with eigenvalue $\lambda_h+\alpha_3\xi^2$. The eigenvalue equation is:
\begin{align}
  v^2\left(\begin{array}{ccc}
                   \frac{\lambda_6}{2}-\frac{1}{2}(\lambda_6-\lambda_5)\xi^2 &\frac{1}{2}(\lambda_6-\lambda_5)s_\beta\,\xi &\frac{1}{2} (\lambda_6-\lambda_5)c_\beta\,\xi \\
                   \frac{1}{2}(\lambda_6-\lambda_5)s_\beta\,\xi & \lambda_H+\epsilon_H\xi^2 & 0 \\
                  \frac{1}{2}(\lambda_6-\lambda_5)c_\beta\,\xi & 0 & \lambda_h+\epsilon_h\xi^2
                  \end{array} \right)\left(\begin{array}{l}
					     a_3\xi\\
					     b_3\xi^2 \\
					     1
					     \end{array}\right)\simeq(\lambda_h+\alpha_3\xi^2)v^2\left(\begin{array}{l}
												  a_3\xi\\
												  b_3\xi^2 \\
												  1
												  \end{array}\right)
\end{align}
From this we get the following:
\begin{align}
 \frac{\lambda_6}{2}\xi\,a_3+&\frac{1}{2}(\lambda_6-\lambda_5)c_\beta\xi\simeq\lambda_h\xi\,a_3 \nonumber \\
 &a_3\simeq\frac{(\lambda_6-\lambda_5)c_\beta}{2\lambda_h-\lambda_6} \\
 \frac{1}{2}(\lambda_6-\lambda_5)&s_\beta\xi^2\,a_3+\lambda_H\xi^2\,b_3 \simeq\lambda_h\xi^2\,b_3 \nonumber \\
 &b_3 \simeq\frac{(\lambda_6-\lambda_5)^2s_\beta c_\beta}{2(2\lambda_h-\lambda_6)(\lambda_h-\lambda_H)}\\
 \lambda_h+\frac{1}{2}(\lambda_6-\lambda_5)&c_\beta\xi^2\,a_3+\epsilon_h\xi^2\simeq \lambda_h+\alpha_3\xi^2 \nonumber \\
 &\alpha_3 \simeq \epsilon_h+\frac{(\lambda_6-\lambda_5)^2c_\beta^2}{2(2\lambda_h-\lambda_6)}
\end{align}
So the lighter \emph{mostly CP-even} state is:
\begin{align}
 \tilde{h}^0 \simeq h^0 + \frac{(\lambda_6-\lambda_5)c_\beta}{2\lambda_h-\lambda_6}\xi\,A^0
\end{align}
with mass 
\begin{align}
 m_{\tilde{h}}^2 \simeq \lambda_h\,v^2 + \epsilon_h\xi^2\,v^2 + \frac{(\lambda_6-\lambda_5)^2c_\beta^2}{2(2\lambda_h-\lambda_6)}\xi^2\,v^2
\end{align}

\end{enumerate}

\providecommand{\href}[2]{#2}\begingroup\raggedright\endgroup

\end{document}